\documentclass{aastex631}

\usepackage[T1]{fontenc}
\usepackage{graphicx}	
\usepackage{amsmath}	
\usepackage{amssymb}	
\usepackage{natbib}
\usepackage[normalem]{ulem}

\usepackage{multirow}
\usepackage{ulem}
\usepackage{bm}

\usepackage{stackengine}

\usepackage{xcolor}

\newcommand{\dotm}{$\dot{m}$}

\newcommand{\ergs}{erg\,s$^{-1}$}

\newcommand{\mergs}{\mathrm{erg\,s}^{-1}}

\newcommand{\Lagr}{\mathcal{L}}

\newcommand{\Jobj}{SDSS J1539+3954}

\graphicspath{{./}{figures/}}

\received{X, 2023}
\revised{X, 2023}
\accepted{X, 2023}

\begin{document}

\title{The disk reverberation mapping of X-ray weak quasars: a case study of SDSS J153913.47+395423.4}

\correspondingauthor{Mouyuan Sun, Jianfeng Wu}
\email{msun88@xmu.edu.cn, wujianfeng@xmu.edu.cn}

\author[0000-0002-1380-1785]{Marcin Marculewicz}
\affiliation{Department of Astronomy, Xiamen University, Xiamen, Fujian 361005, People’s Republic of China}

\author[0000-0002-0771-2153]{Mouyuan Sun}
\affiliation{Department of Astronomy, Xiamen University, Xiamen, Fujian 361005, People’s Republic of China}

\author[0000-0001-7349-4695]{Jianfeng Wu}
\affiliation{Department of Astronomy, Xiamen University, Xiamen, Fujian 361005, People’s Republic of China}

\author[0000-0002-2419-6875]{Zhixiang Zhang}
\affiliation{Department of Astronomy, Xiamen University, Xiamen, Fujian 361005, People’s Republic of China}


\begin{abstract}

The widely adopted ``lamppost'' thermal reprocessing model, in which the variable 
UV/optical emission is a result of the accretion disk reprocessing of the highly 
fluctuating X-ray emission, can be tested by measuring inter-band time lags 
in quasars spanning a range of X-ray power. This work reports the inter-band time 
lag in an apparently X-ray weak quasar, SDSS J153913.47+395423.4. A 
significant cross-correlation with a time delay of $\sim 33$ days (observed-frame) 
is detected in the Zwicky Transient Facility (ZTF) $g$ and $r$ light curves of 
SDSS J153913.47+395423.4. The observed X-ray power seems to be too weak to account for the observed 
inter-band cross-correlation with time delay. Hence the X-ray 
weak quasar SDSS J153913.47+395423.4 is either intrinsically X-ray normal (but observationally X-ray weak), or the 
X-ray emission is not the only mechanism to drive UV/optical variability. In the 
former case, the required X-ray power is at least 19 times stronger than observed, which 
requires either an exceptionally anisotropic corona or Compton-thick obscuration. Alternatively, 
the Corona-heated Accretion disk Reprocessing (CHAR) or the EUV torus models 
may account for the observed time lags.

\end{abstract}

\keywords{Supermassive black holes (1663); Quasars (1319); Active galactic nuclei (16); Accretion (14)} 




\section{Introduction} \label{sec:intro}

The central engine of Active Galactic Nuclei (AGNs), which are powered by accretion of gas onto supermassive black holes (SMBHs), generally is too compact to be spatially resolved and can only be probed in the time or spectral domains. One popular method is disk reverberation mapping, which measures the inter-band time delays of AGN disk continua. The standard thin might be produced, even if it contains incorrect output. Often, compilation errors may be accretion disk model predicts that the emission-region size scales as $R\propto \lambda^{4/3}$ \citep{SS73}, where $R$ is the characteristic radius of the emission at a given wavelength ($\lambda$). The observed inter-band cross-correlation indicates that the same physical mechanism likely drives the variations in different emission regions. The observed time delays are from hours to days \citep[e.g.,][]{Fausnaugh2016RM_NGC5548, McHardy2018, Cackett20,Cackett21RMreview,Guo2022_ZTF_arxiv,GuoWJ22_ADsize_ZTF}. Hence, the time delays are not connected with the radial viscous propagation timescale (i.e., the viscous timescale of a geometrically-thin disk) since this timescale in UV/optical emission regions is longer than $\sim 100$ years. Instead, the radial propagation velocity of the driving mechanism should be close to the speed of light \citep{Krolik91}. One possible driving mechanism is ``lamppost'' thermal reprocessing \citep[e.g.,][]{Cackett2007reprocessing_model}. In this model, the X-ray corona acts as a lamppost and illuminates the accretion disk. The disk absorbs a fraction of the illuminated fluctuating X-ray power and reprocesses it as variable UV/optical thermal emission. Hence, the measured inter-band time delays are the light travel time to various emission regions. 

The disk reverberation-mapping method can test the ``lamppost'' thermal reprocessing model. Indeed, inter-band time delays have been measured for some AGNs spanning a range of luminosity or black-hole mass \citep[e.g.,][]{Fausnaugh2016RM_NGC5548, McHardy2018, Homayouni2019, Guo2022_ZTF_arxiv, GuoWJ22_ADsize_ZTF}. Interestingly, the time delays are not entirely consistent with the X-ray reprocessing of the Shakura \& Sunyaev disk \citep{SS73}. 

There are several studies that have tackled the "time lag discrepancy issue". For instance, disk winds can flatten the disk temperature profile \citep{Sun2019winds} or create a rimmed and rippled accretion disk \citep{Starkey2023}; the disk emission is non-blackbody \citep{Hall2018}; there is a diffuse continuum from extended broad-line regions \citep[e.g.,][]{Cackett18, Korista2019}; coronae can have large scale heights \citep{Kammoun2021}; time lags are not the light travel time \citep{Cai2018, MSun2020_CHAR_melody}.

\begin{figure}[ht]
\centering
    \includegraphics[width = 1.0 \textwidth]{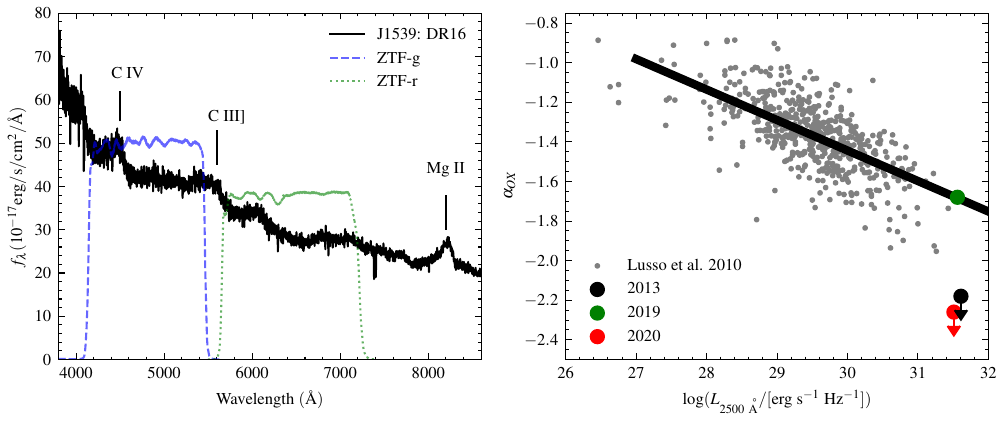}
    \caption{Left panel: The SDSS DR16 spectrum (black curve) of \Jobj\ and the ZTF $g$ and $r$ filter response curves, in blue and green, respectively. 
    Major broad emission lines, i.e., \ion{C}{4}, \ion{C}{3}], and \ion{Mg}{2}, are marked. 
    \ion{C}{4} and \ion{C}{3}] are exceptionally weak in \Jobj. 
    Right panel: The X-ray to UV power-law slope $\alpha_{\mathrm{OX}}$ vs. the rest-frame monochromatic luminosity $L_{2500}$. Grey dots are X-ray-selected type 1 AGN samples by \citet{Lusso2010}. The solid line represents the best-fitting line obtained by \citet{Lusso2010}. The green point represents the X-ray normal state of our target in 2019, well consistent with the relation of \citet{Lusso2010}. Red and black points are the 90\% confidence upper limits for our X-ray weak quasar \Jobj\ \citep{Ni2020_J1539_discovery_of_ex_Xflux}. Its 2013 and 2020 data points are slightly horizontally shifted for better visibility. \Jobj has a significantly lower value of $\alpha_{\rm OX}$ in 2013 and 2020 \citep{Ni2020_J1539_discovery_of_ex_Xflux} than in typical type 1 sources.}
    \label{fig:alpha_OX}
\end{figure}

From the theoretical point of view, the ``lamppost'' thermal reprocessing model has a long-standing energy budget problem \citep[e.g.,][]{Clavel92, Dexter19}. It is well known that the ratio of the X-ray emission at 2~keV to the UV emission at $2500$ \AA\ decreases with increasing UV luminosity (see, e.g., \citealt{Just2007, Lusso2010}). Hence, while the X-ray illumination might be powerful enough to drive the observed UV/optical variability in low-luminosity AGNs, the same process should be inefficient in luminous quasars. The X-ray weakness of a quasar can be described by the relative power of the X-ray and UV emission $\alpha_{\mathrm{OX}}$, which is the ratio of the monochromatic flux density at rest-frame 2 keV to that at rest-frame $2500$ \AA, i.e., 
$\alpha_{\mathrm{OX}}= $ 0.3838 log$(L_{\mathrm{2keV}}/L_{2500\text{\AA}})$ \citep{Tananbaum1979}. The relation between $\alpha_{\mathrm{OX}}$ and $L_{2500}$ \citep[e.g.,][]{Just2007, Lusso2010, Timlin20} can be seen in Fig. \ref{fig:alpha_OX}. For normal AGNs, $\alpha_{\mathrm{OX}}$ is highly correlated with $L_{2500}$ (see the right panel of Fig. \ref{fig:alpha_OX}). For very luminous quasars, the X-ray emission can be much weaker than the UV/optical one. The same issue has been found in 
Seyferts by previous works \citep[e.g.,][]{Uttley2003, Arevalo2008} which showed that, on longer timescales, the optical variability is too large to be driven by the thermal reprocessing.

The energy-budget problem is even more challenging in X-ray weak quasars, a special class of quasars that produce much less X-ray emission than expected from the $\alpha_{\mathrm{ox}}$--$L_{2500}$ relation. A useful parameter is $\Delta \alpha_{\mathrm{OX}}$, which measures the difference between the observed $\alpha_{\mathrm{OX}}$ and expected $\alpha_{\mathrm{OX}}$ from the $\alpha_{\mathrm{OX}}$ - $L_{2500}$ relation \citep[e.g.,][]{Gibson2008, Wu12}. 
At rest-frame 2 keV, $\Delta \alpha_{\mathrm{OX}} = -0.38 $ corresponds to an X-ray weakness of factor 10. 
X-ray weak quasars can have $\Delta \alpha_{\mathrm{OX}}$ of $\sim -0.43$ \citep[e.g.,][]{Ni18,Ni2022J1539_ex_var}, and even up to $-0.81$ for PHL 1811 \citep{Wang2022_PG1001_PG1254}.
The remarkable population of weak-line quasars has a typical equivalent width value of the \ion{C}{4} emission line $\leq 10$ \AA\ \citep[e.g.][]{DS09, Wu12}. Nearly half of the weak-line quasars are also X-ray weak \citep[e.g.][]{Wu11, Wu12, Ni18, Marlar18}. 

There are two physical explanations for their X-ray weakness. First, the X-ray weakness is caused by the intrinsic properties of the corona, which is unable to generate X-ray emission efficiently \citep[e.g.][]{Miniutti2012}. Second, the apparent X-ray weakness is caused by strong gas obscuration due to some "shielding" gas \cite[e.g., the accretion disk, disk wind, and the broad line region (BLR) clouds; see, e.g.,][]{Murray1995, Proga2000, Wu11,Wu12}. Recent results amplify the idea of external absorption caused by winds \citep{Wang2022_PG1001_PG1254} that prevent the observer and the BLR from receiving the X-ray emission. For instance, according to the measured X-ray stacked analysis of X-ray weak quasars \citep{Luo15, Ni18, Ni2022J1539_ex_var} the observed X-ray weakness may have originated in heavy absorption due to small-scale shielding. The hard X-ray spectra show average effective power-law photon indices of $\sim$ 1.2--1.4. These values suggest high levels of intrinsic X-ray absorption of $N_{\rm H} \geq 10^{23}\ \rm{cm^{-2}}$. Nevertheless, the observed X-ray luminosity (which can significantly underestimate the intrinsic X-ray power) is too small to drive the observed UV/optical variability. The sight line from the corona to the disk might not be heavily obscured; it is unclear whether the illuminating X-ray is able to induce the observed variations in disk emission or not. Hence, it is interesting to explore the UV/optical cross-correlations in X-ray weak quasars. 

So far, the disk mapping is performed for AGNs with low luminosity (thus strong X-ray power) or X-ray normal quasars. If the ``lamppost'' thermal reprocessing is correct, we expect no inter-band cross-correlations with time delays in the UV/optical light curves of X-ray weak quasars. Hence, X-ray weak quasars are ideal for verifying the X-ray reprocessing and probing alternative possible UV/optical variability mechanisms. Here we report the time delay of the $g$ and $r$ band continua of an X-ray weak quasar, SDSS J153913.47+395423.4 (hereafter \Jobj). To our best knowledge, \Jobj\ has the highest black-hole mass, optical luminosity, and redshift among AGNs with continuum time lag measurements. This manuscript is formatted as follows. In Section \ref{sec:J1539_properties}, we describe the properties of the X-ray weak quasar \Jobj. In Section \ref{sec:results}, we present our results. In Section \ref{sec:discussion}, we discuss the physical implications of our results. A summary is given in Section \ref{Sec:conclusion}.

\section{\Jobj\ properties}\label{sec:J1539_properties}

\Jobj\ lies at redshift $1.935$, with J2000 coordinates RA: 234.806121$^{\circ}$, DEC: +39.906525$^{\circ}$, and an $i$-band apparent magnitude $i=17.38$ \citep{SDSSDR4}. The black-hole mass $\log (M_{\mathrm{BH}}/M_{\odot})=9.8$ with the typical uncertainty of 0.4 dex, and the bolometric luminosity $\log (L_{\mathrm{bol}}/[\mergs])=47.3207\pm 0.0009$ \citep{WuShen22} estimated based on the optical luminosity at rest-frame 3000 \AA. Note that this black-hole mass is estimated via the broad \ion{Mg}{2} line. We caution, however, that there are caveats associated with the \ion{Mg}{2} line-based virial black-hole mass, such as profile complexity, profile shape, spectral noise \citep[e.g.,][]{Marziani2013, Sun_Mg_2015, Kovacevic2017}. On the left panel of Fig. \ref{fig:alpha_OX}, we show the SDSS DR16 spectrum \citep{SDSSDR16} of \Jobj\ and the ZTF $g$ and $r$ filter response curves. Major broad emission lines, e.g., \ion{C}{4}, \ion{C}{3}], and \ion{Mg}{2} are marked in the figure. Due to its large redshift, the more reliable H$\beta$ mass estimator is not covered by the SDSS spectrum.

According to the observed-frame $0.5$--$2$ keV fluxes \citep{Luo15,Ni2020_J1539_discovery_of_ex_Xflux,Ni2022J1539_ex_var}, the corresponding Galactic absorption-corrected $2$--$10$ keV X-ray luminosities of \Jobj\ are $\log(L_{X}/[\mergs]) \leq 43.76$, = $45.04$, and $<44.07$ in the 2013, 2019, and 2020 X-ray observations, respectively. Even during the X-ray flare stage (in 2019), \Jobj\ shows a negligible amount of X-ray power compared to its UV/optical luminosity. \Jobj\ shows strong variations in $\alpha_{\mathrm{OX}}$ \citep{Luo15,Ni2020_J1539_discovery_of_ex_Xflux,Ni2022J1539_ex_var}. During nearly 7 years, its value went from < $-$2.18 to $-$1.68 and back to < $-$2.26 (X-ray weak) in 2013, 2019, and 2020, respectively (see Fig.~\ref{fig:alpha_OX}). \cite{Ni2020_J1539_discovery_of_ex_Xflux} studied the extreme X-ray variability in \Jobj\ and proposed its nature as being due to a geometrically and optically thick accretion disk whose thickness varies.

\begin{figure}[ht]
    \centering
    \includegraphics[width = 0.8\textwidth]{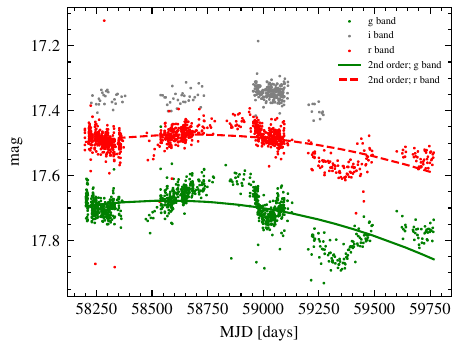}
    \caption{The \textit{g}, \textit{r}, and \textit{i}-band light curves of the \Jobj\ from ZTF. The green solid and red dashed curves are the best-fitting second-order polynomial curves for \textit{g} and \textit{r}, respectively. Due to the limited coverage of the $i$ band, we will not analyze the $i$-band light curves. For the sake of completeness, we have included it in the plot }.
    \label{fig:J1539_dr13_all}
\end{figure}

We examine \Jobj\ in two bands ($g$ and $r$) using the ZTF data. We retrieve the ZTF $g$, $r$, and $i$ band light curves of \Jobj\ via the IRSA GATOR ZTF catalog service\footnote{https://irsa.ipac.caltech.edu/cgi-bin/Gator}. ZTF is a time-domain survey with the 1.2 meters Schmidt telescope at the Palomar Observatory \citep{ZTF_data_processing2019}. Each ZTF exposure time is 30 seconds with a $5\sigma$ sensitivity limit of $20.8$ and $20.6$ in the $g$ and $r$ bands, respectively. We use the point spread function (PSF) magnitudes with the "catflags" parameter equal to zero \citep[i.e., good quality data, not affected by clouds and/or the moon, or instrument; see][]{ZTF_data_processing2019}. The total time duration of the light curves are: 1564, 1568, and 1043 days, for $g$, $r$, and $i$ bands, respectively. The median cadence of each light curve is: 1.50, 0.79, and 3.52 days for $g$, $r$, and $i$ bands, respectively. The  ZTF's $g$, $r$, and $i$ bands light curves, with the best-fitting 2nd order polynomial curves for $g$ and $r$, are presented in Figure \ref{fig:J1539_dr13_all}. The $g$ and $r$ light curves show prominent variations on short and long timescales. The $i$-band light curve is sparse and is excluded in the subsequent time lag analysis. 

\section{Results} \label{sec:results}
We measure the cross-correlation and time delay of the ZTF $g$ and $r$ light curves of \Jobj\ (Fig. \ref{fig:J1539_dr13_all}). To determine the inter-band time lags, we use PyCCF \citep{PyCCF_MSun_2018}, a Python version of the interpolation cross-correlation function code based on \cite{Peterson1998PASP}. The light curves used in the PyCCF are from MJD = 58200 days to MJD = 59800 days. The time delay ($\tau$) searching range was from $-$200 days to 200 days with a 0.2-day step. The PyCCF code uses the flux randomization/random-subset selection method to estimate the time lag uncertainties.

\begin{figure}
    \centering
    \includegraphics[width= 0.7 \textwidth]{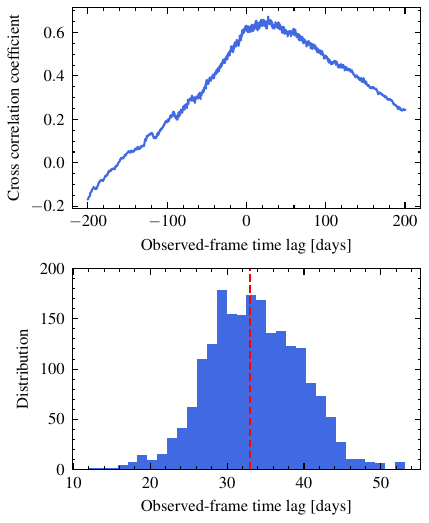}
    \caption{Top panel: The observed-frame time lag between $g$ and $r$ bands in days vs. correlation coefficient for the 2nd-order polynomial approach. This approach is based on subtracting the 2nd-order polynomial linear least-squares fit (with equal weight given to all data points) from the original light curves. Bottom panel: The $g$-vs-$r$ time delay ($\tau_{\mathrm{gr, obs}}$) centroid distribution (from the flux randomization/random-subset selection method) for the 2nd-order polynomial approach. The red dashed line indicates the median value.}
    \label{fig:200d_time_lag_distrib}
\end{figure}

The cross-correlation analysis can be biased if the long-term trend is not adequately subtracted from the data \citep{Welsh1999}. Following previous studies \citep[e.g.,][]{Fausnaugh2016RM_NGC5548}, we adopt the following two steps to detrend the ZTF light curves. First, we fit low-order polynomial functions to the observed light curves via the least squares method. Second, we subtract the best-fitting polynomial functions from the observed light curves. The residuals are used for subsequent cross-correlation analysis. We use both the first and second-order polynomial functions (see curves in Fig.~\ref{fig:J1539_dr13_all}). The cross-correlation function of the second-order polynomial approach and the corresponding distribution (obtained via the flux randomization/random-subset selection method) for the time lag between $g$ and $r$ bands ($\tau_{\mathrm{gr, obs}}$) in the observed-frame are presented in Fig.~\ref{fig:200d_time_lag_distrib}. The observed-frame time delays and their $1\sigma$ uncertainties are $40^{+6}_{-5}$ days and $33^{+6}_{-5}$ days, respectively. The two-time delays are statistically consistent with each other within their $1\sigma$ uncertainties (i.e., $\simeq 5$ days). Note that the time delay of the observed ZTF light curves without detrending is $36 \pm 5$ days (observed-frame), which also statistically agrees with our first and second-order polynomial results. In summary, a significant cross-correlation with a time delay of $\tau_{\mathrm{gr, obs}} \sim 33$ days (observed-frame) is detected in the X-ray weak quasar \Jobj. 

The theoretical time delay can be calculated for the X-ray reprocessing of the Shakura \& Sunyaev disk. We use the black-hole mass and bolometric luminosity from Section \ref{sec:J1539_properties}. Following Eq. 1 in \citet{Li2021}, we calculate the expected $\lambda$ = 2500 \AA\ disk lag ($\tau_{2500, \mathrm{th}}$) with respect to the driving light curve, i.e., $\tau_{2500, \mathrm{th}}=15.8$ days (rest-frame). The theoretical rest-frame time delay between the $g$ and $r$ bands is $\tau_{\mathrm{gr, th}}=\tau_{2500, \mathrm{th}}((\lambda_r/2500)^{4/3}-(\lambda_g/2500)^{4/3})=3.8$ days, where $\lambda_g=4776/(1+z)$ and $\lambda_r=6231/(1+z)$ are the rest-frame wavelengths of the $g$ and $r$ bands, respectively. Hence, the theoretical time lag in the observed-frame is $\tau_{\mathrm{gr, th}}(1+z)=11.2$ days, which is smaller than the observed one ($\tau_{\mathrm{gr, obs}} \sim 33$ days) by a factor of three. This larger-than-expected time lag result is observed in many previous studies \citep[e.g.,][]{Fausnaugh2016RM_NGC5548, Cackett18, Edelson19,Netzer22, Guo2022_ZTF_arxiv, GuoWJ22_ADsize_ZTF}. 

Given its large $M_{\mathrm{BH}}$, the observed $g$ and $r$ emission regions of \Jobj\ are, unlike lower-mass Seyferts, very close to the SMBH. Indeed, according to the Shakura \& Sunyaev disk model, the average sizes of the $g$ and $r$ emission regions are $c\tau_{2500, \mathrm{th}}(\lambda_g/2500)^{4/3}=24\ R_g$ and $c\tau_{2500, \mathrm{th}}(\lambda_r/2500)^{4/3}=35\ R_g$, respectively, where $R_g=GM_{\mathrm{BH}}/c^2$. Hence, general-relativistic corrections and inner-boundary effects can be significant when estimating the theoretical time delay. These effects are considered by \cite{Kammoun2021_an_pres} and \cite{Kammoun2021}; they provide an analytic prescription of the expected time delay. Based on \cite{Kammoun2021_an_pres}, we calculated the $g-r$ time lag for our target with the following parameters: $\log (M_{\mathrm{BH}}/M_{\odot})=9.8$, the dimensionless accretion rate \dotm$ = 0.26$ \citep{WuShen22}, X-ray luminosity $\log(L_{X}/[\mergs]) = 45.04$ \citep{Ni2020_J1539_discovery_of_ex_Xflux} at normal state, and the corona height is $10$ $R_g$. Several studies \citep[e.g.,][]{Kammoun2021} adopt much higher corona heights (e.g., $\sim 100$ $R_g$) for some sources; in our opinion, the corona height in \Jobj\ cannot be large; otherwise, the disk would not effectively obscure the X-ray emission. The $g-r$ time lag for the corona height $= 10$ $R_g$ is $17.19$ days in the rest-frame. Recently, a new analytical prescription that involves disk color correction is also obtained (priv. comm., Kammoun 2023 et al. in prep.); with a color correction factor of $1.6$, the theoretical $g-r$ time lag is (rest-frame) $10.37$ days. These results are roughly consistent with our observations. 

Note that the virial black-hole mass might be underestimated in weak-line quasars \citep[e.g.,][]{MM2020} and \Jobj\ is also a weak-line quasar \citep[e.g.,][]{Wu11, Wu12, Plotkin15, Luo15}. If so, the theoretical time lag is also underestimated. We will discuss this in Section~\ref{sec:discussion}. 

\section{Discussion and Physical implications}\label{sec:discussion}
\subsection{The driving mechanism for the UV/optical variability}
Our results (i.e., evident cross-correlations with time lags) for \Jobj\ are unexpected in the X-ray reprocessing model and have important implications for our understanding of the structure of accretion disks and the interplay between accretion disks and coronae. To explain the results, we propose several possibilities.

If the ``lamppost'' thermal reprocessing is correct, the X-ray luminosity that illuminates the accretion disk should be much stronger than we observed. 
On 2013-12-13, the 2-10 keV X-ray luminosity was $<$$ 5.72\times 10^{43}$ \ergs. In 2019-09-12 it increased to $1.09\times 10^{45}$ \ergs\ and in 2020-06-17 it went back to the low stage of <$1.17\times 10^{44}$ \ergs \citep{Ni2022J1539_ex_var}. Since the observed UV/optical variability amplitude is about 10\%, the required X-ray luminosity should be at least 10\% of the UV/optical luminosity. In the ``lamppost'' thermal reprocessing model, the whole disk emission will vary in response to the X-ray. Suppose we assume the variability amplitudes of the disk emission in other bands (which is not probed by ZTF) are also around $10\%$. We can conclude that the required X-ray power should be at least $10\%$ of the bolometric luminosity. The bolometric luminosity is $2.09\times10^{47}$ \ergs\ which indicates that the required X-ray power is $\sim 2.09\times10^{46}$ \ergs. Hence, the required X-ray power is at least 19 times larger than the observed one (2019). An even higher value of X-ray power (a factor of $\geq 178$) is required for the 2020 X-ray observation.

We present two possible solutions: i) the X-ray emission is extremely anisotropic. ii) the ambient gas may act as a ``shielding'' gas and blocks most of the X-ray photons. Hence, the intrinsic corona is strong, but the observed X-ray weakness could be caused by dust-free gas absorption \citep{Wang2022_PG1001_PG1254}, which might be related to the disk winds. Note that the X-ray spectrum of the \textit{CHANDRA} observation in 2019 has a typical quasar photon index of $\Gamma=2$ \citep{Ni2020_J1539_discovery_of_ex_Xflux}, i.e., no significant intrinsic absorption. Hence, the observed X-ray luminosity is the same as the intrinsic one. Even in this case, the X-ray emission is not powerful enough to drive the variability in the disk emission. If the ``lamppost'' thermal reprocessing model is incorrect, some alternative explanations are proposed (e.g., \citealt{Li2021}, and references therein). First, we consider the disk-corona magnetic coupling scenario, also known as the CHAR model \citep{MSun2020_CHAR_melody}. In this model, strong magnetic fluctuations in the corona rather than the X-ray illumination induce the variations in the disk and produce the multi-band correlated variations with time delays (for a more detailed discussion, please see \citealt{MSun2020_CHAR_melody}). The corresponding time lags are not light-travel time delays. For luminous quasars like J1539, the time delay of the CHAR model without general-relativistic effects is similar to $\tau_{gr,\mathrm{th}}$ \citep[see][]{Li2021}. However, we caution that the general-relativistic CHAR model is still under development and a detailed comparison to our results is beyond the scope of this work. Second, our results can be understood in the EUV reprocessing model. In this model, it is the EUV emission coming from the innermost "EUV region" rather than the X-ray that illuminates the outer disk \citep{Gardner17} and drives the UV/optical variability. A puffed thick disk, which is widely proposed to form in X-ray weak quasars, might act as the "EUV torus". Future determination of the rest-frame UV/optical time lag-wavelength relation of \Jobj\ (via, e.g., LCOGT) can test these mechanisms. 

\subsection{The scaling relation between inter-band time lags and luminosities}
Study of \Jobj\ allows us to extend the continuum emission radius-luminosity relation at 5100 \AA\ to the high-luminosity end. In fact, \Jobj\ is almost one order of magnitude more luminous than the most luminous quasars in previous time lag-luminosity studies \citep{Guo2022_ZTF_arxiv, Netzer22, Montano22}. To obtain the radius at 5100 \AA\ ($R^{\mathrm{rest}}_{5100}$) from the time lag, we assume that $R_{\lambda} \propto \lambda^{4/3}$ \citep{Guo2022_ZTF_arxiv}. Hence, we expect that 

\begin{equation}
      R^{\mathrm{rest}}_{5100} = \frac{\tau_{\mathrm{gr,obs}}/(1+z)} {\left(\lambda_r/5100\right)^{4/3} - \left(\lambda_g/5100\right)^{4/3}} = 120 \pm 18\ \mathrm{(lt-days)} \\,
     \label{eq:R5100_RF}
\end{equation}
where $\tau_{gr, obs}$, $\lambda_g$, and $\lambda_r$ are defined in Sec. \ref{sec:results}. We also include the ``core'' sample and local AGNs of \cite{Guo2022_ZTF_arxiv} for subsequent analysis. 

To estimate $L_{5100}$ of \Jobj, we use the same $L_{\mathrm{bol}}$ as in Sec.~\ref{sec:J1539_properties} \citep[which was estimated from $L_{3000}$; see Sec. 4.1 of][]{WuShen22}, and the $L_{5100}$ bolometric factor of $10$ \citep{Richards2006}, i.e., $L_{5100}=L_{\mathrm{bol}}/10$. Following \cite{Guo2022_ZTF_arxiv}, we assumed the same uncertainty, i.e.,  0.05 dex. Hence, 
$\log (L_{\mathrm{5100}}/[\mergs])=46.32\pm 0.05$.

We assume the scaling relation between $R_{5100}$ and $L_{5100}$ follows
\begin{equation}
\label{eq:rl}
    \log (R_{5100}/[\mathrm{lt-days}])=m\log(L_{5100}/[\mathrm{erg\ s^{-1}}])+b \\,
\end{equation}
where $m$ and $b$ are the slope and intercept of the relation. We fit Eq.~\ref{eq:rl} to SDSS J1539+3954, the ``core'' sample, and the local AGN sample in \cite{Guo2022_ZTF_arxiv}. We adopt the maximum likelihood approach and use MCMC (via \texttt{emcee}) to sample the following likelihood \citep{Hogg2010stat},
\begin{equation}
    \Lagr = - \frac{1}{2}\sum_n\left[ \frac{(\log (R^{\mathrm{rest}}_{5100}/[\mathrm{lt-days}]) - m\log(L_{5100}/[\mathrm{erg\ s^{-1}}])-b )^2}{s^2_n}
    + \mathrm{ln}(2\pi s^2_n) \right]\\,
\end{equation}
where
\begin{equation}
    s^2_n = (\delta \log (R^{\mathrm{rest}}_{5100}/[\mathrm{lt-days}]))^2 + f^2 (m\log(L_{5100}/[\mathrm{erg\ s^{-1}}])+b)^2 \\,
\end{equation}
and $\delta \log (R^{\mathrm{rest}}_{5100}/[\mathrm{lt-days}])$ and $f$ are the measurement error of $\log (R^{\mathrm{rest}}_{5100}/[\mathrm{lt-days}])$ and the intrinsic scatter factor, respectively. 

\begin{figure}[ht]
\centering
    \includegraphics[width = 0.8 \textwidth]{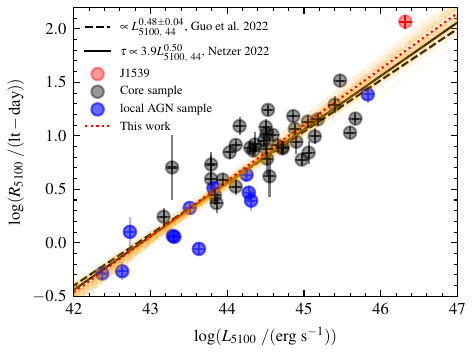}
    \caption{The radius-luminosity relation at 5100 \AA. The best linear fit (red dotted line) applies to all sources with a typical luminosity uncertainty of 0.05 dex. 
    The black solid and dashed lines correspond to the relations in \cite{ Netzer22} and \cite{Guo2022_ZTF_arxiv}. \Jobj\ helps to extend the radius-luminosity relation to the higher end of $L_{5100}$. Given its weak-line nature, one would expect that \Jobj\ falls below the radius-luminosity relation if the diffuse continuum is the only mechanism that accounted for this relation. This expectation is inconsistent with our observation.}
    \label{fig:GN_J15} 
\end{figure}

From the MCMC chains, we find the median values of $m$ and $b$ and their $1\sigma$ uncertainties. Our fitting results are $m=0.52\pm0.03$ and $b=-22^{+1.4}_{-1.4}$. The best-fitting result, which is shown in Figure~\ref{fig:GN_J15}, is statistically consistent with \cite{Guo2022_ZTF_arxiv} and \cite{Netzer22}. Note that our target is not included in their fitting process. 

Unlike other sources, SDSS J1539+3954 is a weak-line quasar. As a result, the diffuse continuum from the inner BLR clouds should also be weak since its emission regions are smaller than the H$\beta$ emission regions by a factor of eight \citep{Netzer22}, i.e., comparable to the emission regions for high-ionization lines. Hence, if the diffuse continuum is responsible for the scaling relation between $R_{5100}$ and $L_{5100}$ \citep[as proposed by][]{Netzer22}, we expect that SDSS J1539+3954 lies below the scaling relation. This expectation is inconsistent with the data. Hence, our results indicate that the diffuse continuum might not be the only mechanism to produce the scaling relation between $R_{5100}$ and $L_{5100}$. 

We now discuss the impacts of the virial black-hole mass estimations. Recently, \cite{MM2020} adopt the continuum fitting method \citep[e.g.,][]{Castignani2013, Capellupo2015} to estimate $M_{\mathrm{BH}}$ in WLQs. Comparing with our SED fitting $M_{\mathrm{BH}}$, the virial black-hole mass based on $\mathrm {H\beta}$ is systematically smaller by an average factor of four to five. If the \ion{Mg}{2}-based virial mass of SDSS J1539+3954 is similarly smaller than the actual $M_{\mathrm{BH}}$ by a factor of $\sim 4$, we expect that the theoretical time lags in Section~\ref{sec:results} should increase by a factor of $\sim 4^{1/3}=1.5$, i.e., this mass discrepancy still cannot account for the difference between the theoretical and observed time lags.

\section{Conclusion}\label{Sec:conclusion}

In this study, we use ZTF observations to explore the inter-band time lag between $g$ and $r$ for an X-ray weak quasar, \Jobj. Our main findings can be summarized as follows. First, a significant cross-correlation with a time delay of $\tau_{\mathrm{gr, obs}}=33^{+6}_{-5}$ days (observed-frame) is detected. Our observed time delay is about three times larger than the X-ray reprocessing in a Shakura \& Sunyaev disk if the virial black-hole mass is valid and if the lag is dominated by a simple light-travel time. The X-ray reprocessing in a general-relativistic disk can account for the observed time lag. Second, to understand the observed time delay, the X-ray power of \Jobj\ should be at least 19 times larger than the X-ray observations. Such inconsistency in the ``lamppost'' thermal reprocessing model is possible if: a. the corona emission is anisotropic; b. strong absorption is present (e.g., by winds). Alternatively, we argue that the magnetic coupling scenario \citep{MSun2020_CHAR_melody} or the EUV reprocessing model can explain our results \citep{Gardner17}. Additionally, due to the high luminosity of \Jobj, we were able to extend the range of the radius-luminosity relation (see Fig.  \ref{fig:GN_J15}). We obtained $R^{\mathrm{rest}}_{5100} = 120 \pm 18\ \mathrm{(lt-days)}$ which lies slightly above the $R^{\mathrm{rest}}_{5100}$-luminosity relation of \cite{Netzer22}. This relation is proposed to be driven by the diffuse continuum from the BLR. The diffuse continuum is rather weak in \Jobj\ since this source is a weak emission-line quasar. Hence, we would expect that \Jobj\ lies below the $R^{\mathrm{rest}}_{5100}$-luminosity relation. The inconsistency between our observation and expectation suggests that other/additional mechanisms (e.g., the CHAR model) could also contribute to the $R^{\mathrm{rest}}_{5100}$-luminosity relation. Further investigation of the inter-band time lags in other X-ray weak AGNs will bring opportunities to understand the physical origin of quasar UV/optical variability, inter-band time lags, and disk sizes. 


\section*{Acknowledgements}

\begin{acknowledgments}
We would like to thank the anonymous referee for useful comments that improved the clarity of the paper. M.M. and M.Y.S. would like to thank Bin Luo for the valuable discussion. M.M. would like to thank Elias Kammoun for the valuable discussion. M.Y.S. acknowledges support from the National Natural Science Foundation of China (NSFC-11973002), the Natural Science Foundation of Fujian Province of China (No.\ 2022J06002), and the China Manned Space Project grants (No.\ CMS-CSST-2021-A06; No.\ CMS-CSST-2021-B11). J.W. acknowledges support from the National Natural Science Foundation of China (NSFC-12273029 and NSFC-12221003), and the China Manned Space Project grants (No.\ CMS-CSST-2021-A05; No.\ CMS-CSST-2021-A06). Z.X.Z. acknowledges support from the National Natural Science Foundation of China (NSFC-12033006 and NSFC-12103041).

Based on observations obtained with the Samuel Oschin Telescope 48-inch and the 60-inch Telescope at the Palomar Observatory as part of the Zwicky Transient Facility project. ZTF is supported by the National Science Foundation under Grants No.\ AST-1440341 and AST-2034437 and a collaboration including current partners Caltech, IPAC, the Weizmann Institute for Science, the Oskar Klein Center at Stockholm University, the University of Maryland, Deutsches Elektronen-Synchrotron and Humboldt University, the TANGO Consortium of Taiwan, the University of Wisconsin at Milwaukee, Trinity College Dublin, Lawrence Livermore National Laboratories, IN2P3, University of Warwick, Ruhr University Bochum, Northwestern University and former partners the University of Washington, Los Alamos National Laboratories, and Lawrence Berkeley National Laboratories. Operations are conducted by COO, IPAC, and UW.
\end{acknowledgments}

%

\vspace{5mm}
\facilities{ZTF \citep{ZTF-DOI}}


 \software{astropy \citep{astropy}, matplotlib \citep{matplotlib}, numpy \citep{numpy}, PyCCF \citep{PyCCF_MSun_2018}, emcee \citep{emcee}.}

\bibliography{J1539.bib}
\bibliographystyle{aasjournal}
\end{document}